\documentclass[12pt,a4paper]{article}
\usepackage{amsmath}
\usepackage{graphicx}
\usepackage{times}
\begin{document}
\begin{titlepage}
\title{Spin measurements in strangeness production at the LHC}
\author{ S.M. Troshin, N.E. Tyurin\\[1ex]
\small  \it SRC IHEP of NRC ``Kurchatov Institute''\\
\small  \it Protvino, 142281, Russian Federation}
\normalsize
\date{}
\maketitle
  \begin{abstract}
 We briefly recollect the problem of significant hyperon polarization emphasizing the role of spin in dynamics of hadron interactions. We provide also some model predictions based on chiral dynamics and the impact parameter picture for  illustration.   
The old,  but yet unsolved problem of hyperon polarization, can obtain a new insight from the measurements at the LHC energies and,  in combination with other measurements, can be  used for tagging QGP formation in $pp$--collisions. 
\end{abstract}
\end{titlepage}
\setcounter{page}{2}
\section*{Introduction}
The importance of spin measurements for studies of hadron--interaction dynamics is well known. However, there are very few well established experimental facts regarding behavior and dependence of spin observables at high energies. In particular,the knowledge of the energy dependence of such observables  is crucial for the analysis  since it could  
justify, neglecting the spin degrees of freedom. The LHC machine provides record energy of collisions and could definitely help in this matter. The experiments at the LHC have discovered several collective effects in such small systems  as $pp$--interactions (cf. reviews \cite{coll,trib} and references therein to the original papers.) 

The  feasible experimental direction of the spin studies is based on the self-analyzing particle decays, the mostly known example of which is the weak decay of $\Lambda$-hyperon (cf. e.g \cite{newrev}). To study  energy dependence of spin effects it is most relevant to choose the respective reaction where measurements have already been performed in the widest range of energy variation, namely, the measurements of the polarization of the above mentioned final $\Lambda$-hyperon in the inclusive reaction $pp\to\Lambda X$.
It is the most interesting and persistent for a long time spin phenomena
which was observed in inclusive hyperon production in collisions of
unpolarized hadron beams. A very significant polarization of
$\Lambda$--hyperons has been discovered more than four  decades ago (cf. \cite{newrev} and references therein) and
experimentally, the process
of $\Lambda$-production has been studied more extensively than other hyperon
production processes.
Therefore, we concentrate on the particular pattern  of $\Lambda$--polarization.
In addition, the spin structure of this particle seems rather simple and
is determined by spin of strange quark  in the $SU(6)$-quark model.

Experimental
results on hyperon polarization are widely known, those are being stable during a long time and well documented.
Polarization of $\Lambda$ produced in the unpolarized inclusive $pp$--collisions
is negative (it is perpendicular to production plane and directed opposite to the normal to this plane) and energy
independent. It increases linearly with $x_F$ at large transverse momenta
($p_\perp\geq 1$ GeV/c),
and for such
values of transverse momenta   is almost
$p_\perp$-independent. The comprehensive review of the experimental situation is given in \cite{newrev}. It should be noted that the above results
are for the hyperons which appear to be the proton's fragmentation products.
No  energy independence of global polarization has been observed for  $\Lambda$'s  produced in the  nuclear $AuAu$--interactions \cite{lisa}.
The recent RHIC data (with high error bars, though) could imply energy decrease of global polarization of $\Lambda$ and could testify that the different
leading dynamical mechanisms result in the strange quark polarization in  nuclear collisions. However, one should also take into account that the data in this case are for the polarization measured with respect to reaction (not production) plane

Perturbative QCD
amended  by the collinear factorization scheme
leads to vanishing values of $\Lambda$--polarization
\cite{pump,gold} and does not correspond to the experimental data values and observed trends.
Including the higher twists contributions allows one
to obtain higher values for polarization but does not change qualitative
 dependence 
$p_\perp^{-1}$ at large transverse momenta\cite{efrem,sterm,koike}.
It is difficult to reconcile the decreasing dependence  with the flat  one observed in the
data.  Inclusion of the parton internal transverse momenta
($k_\perp$--effects) into the
polarizing fragmentation functions  leads also   to decreasing trend of polarization \cite{anselm}. It allows again to change the scale only.

The aim of this  note is to provide arguments for measurements of the hyperon polarization at the LHC energies. 
For that purpose we discuss the mechanism leading to a nondecreasing energy
dependence of the  transverse polarization of $\Lambda$ produced in $pp$-collisions.
We also briefly present an experimental feasibility of such measurements and point out to the role of polarization studies in the strangeness production as a complementary tool for QGP detection in small systems..

\section{Mechanism of the strange quark polarization}
In the simple quark model the u- and d-quarks in $\Lambda$ are coupled
to S=0, I=0 diquark. It is strange quark polarization is responsible for the significant transverse polarization of $\Lambda$. Thus, to explain the 
$\Lambda$-polarization the dynamical mechanism of the strange quark polarization should be developed.
There were several proposals for an explanation of the  polarization
of strange quarks produced in the collisions of the unpolarized protons. Among them one should note the mechanism based on Thomas precession  \cite{thom} and Lund model  \cite{lund}. The both explanations are semiclassical in its nature. There is another semiclassical mechanism
based on chiral spin filtering. We briefly mention it in order to stress that the measurements of the final hyperon polarization  at the LHC energies could have sense since this particular mechanism leads to a nonvanishing polarization of $\Lambda$ when collision energy increases.

The polarization of the strange quarks  might happen to originate from a genuine 
nonperturbative  QCD (cf. e.g. \cite{spin02}). 

In the nonperturbative sector of  QCD, there are two important
phenomena,  confinement and spontaneous breaking of chiral symmetry ($\chi$SB). 
The  relevant scales   are defined by the
parameters $\Lambda _{QCD}$ and $\Lambda _\chi $.  
Chiral $SU(3)_L\times
SU(3)_R$ symmetry is spontaneously  broken  at the distances which are
in  the range between
the above two scales.  The $\chi$SB mechanism leads
to generation of quark masses and appearance of quark condensates. It describes
transition of current into  constituent quarks.
  Constituent quarks are considered to be the quasiparticles, i.e. they
are a coherent superposition of bare  quarks and their masses
have a magnitude comparable to  a hadron mass scale.  The
hadron  is  represented as a loosely bounded system of the
constituent quarks.
The well known direct result of the $\chi$SB mechanism   is  appearance of the Goldstone bosons (GB).

The particles (protons) in the initial state   are unpolarized.
Absence of polarization means that  probabilities of states with spin up and spin down are equal.
The main idea of the  mechanism is   filtering
of the two initial spin states with due to different strength of interactions. 
This filtering acts like polaroid and leads to polarization of the particles in final state ($\Lambda$, in particular).
The specific
mechanism of such filtering can be developed on the basis of chiral quark model.
Namely, we exploit the feature of chiral quark model that constituent quark $Q_\uparrow$
with transverse spin in up-direction can fluctuate into Goldstone boson and
  another constituent quark $Q'_\downarrow$ with opposite spin direction
performing a spin-flip transition \cite{cheng}:
\begin{equation}\label{trans}
Q_\uparrow\to GB+Q'_\downarrow\to Q+\bar Q'+Q'_\downarrow.
\end{equation}
To compensate quark spin flip $\delta {\bf S}$ an orbital angular momentum
$\delta {\bf L}=-\delta {\bf S}$ should be generated in the final state of reaction (\ref{trans}).
The presence of this orbital momentum $\delta {\bf L}$  in its turn
means a certain shift in the impact parameter
value of the final quark $Q'_\downarrow$ (which in its turn is transmitted to the shift in the impact
parameter of $\Lambda$)
\[
\delta {\bf S}\Rightarrow\delta {\bf L}\Rightarrow\delta{\bf b}_{Q'}.
\]
Due to   different strengths of interaction at the different values of the
impact parameter, the processes of transition to the
spin up and down states will have different probabilities which  leads eventually to
polarization of $\Lambda$.

\begin{figure}[h]
\begin{center}
  \resizebox{4cm}{!}{\includegraphics*{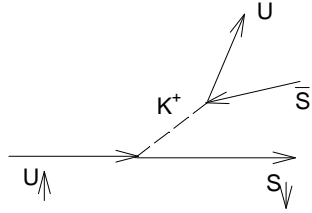}}
\end{center}
\caption{Transition of the spin-up constituent quark $U$ to the spin-down strange quark.
 \label{ts1}}
\end{figure}
In the  case of $\Lambda$--polarization, the relevant transitions
of constituent quark $U$ (cf. Fig. 1) is correlated with the shifts $\delta b_S$
in impact parameter $b_S$ of the final
strange quark, i.e.:
\begin{eqnarray}
  \nonumber U_\uparrow & \to & K^+ + S_\downarrow\Rightarrow\;\;-\delta{\bf b}_S \\
\label{spinflip} U_\downarrow & \to & K^+ + S_\uparrow\Rightarrow\;\;+\delta{\bf b}_S.
\end{eqnarray}
Relations (\ref{spinflip}) clarify the mechanism  of spin states filtering:
 i.e. when shift in impact
parameter is $-\delta{\bf b}_S$ the
interaction is stronger compared to the case when shift is $+\delta{\bf b}_S$,
and the final $S$-quark
(and $\Lambda$-hyperon) becomes negatively polarized. The adopted mechanism of the spin states filtering is associated with the
 emission of Goldstone bosons by the constituent quarks.

It is important to note here that the shift of ${\bf b}_\Lambda$
(the impact parameter of the final hyperon)
is correlated with the shift of the impact parameter of the initial particles according
to the relation between impact parameters in the multiparticle production \cite{webb}:
\begin{equation}\label{bi}
{\bf b}=\sum_i x_i{ {\bf  b}_i}.
\end{equation}
Let the variable $ b_\Lambda$ to be conjugated to the transverse momentum of $\Lambda$,
but relations  between functions depending on the impact parameters
$b_i$ are nonlinear.
Since we are considering production of $\Lambda$ in the fragmentation region, (i.e.
at large $x_F$)  the following approximate relation
\begin{equation}\label{bx}
b\simeq x_F b_\Lambda,
\end{equation}
which results from Eq. (\ref{bi})\footnote{With an assumption on the
smallness of Feynman $x_F$ for other secondary particles.} is adopted.

The main interest of this note is a trend for 
the polarization energy dependence, which can be checked experimentally. 
We note only that $\delta b_S$ (we assume that $\delta b_S\simeq \delta b_\Lambda$) can be connected with the radius of quark interaction
$r_{U\to S}^{flip}$
responsible for the transition $U_\uparrow\to S_\downarrow$ changing quark spin and flavor:
\[
\delta b_S\simeq r_{U\to S}^{flip}.
\]

To evaluate polarization dependence on $x_F$ and $p_\perp$
we use semiclassical correspondence  between transverse momentum and impact parameter
  values.
 The energy and $p_\perp$-independent behavior
of polarization $P_\Lambda$ takes place at large values of $p_\perp$:
\begin{equation} P_\Lambda(s,\xi)\propto -x_Fr_{U\to S}^{flip}
{M}/\zeta.\label{polllg}
\end{equation}
This flat transverse momentum dependence results from the similar
rescattering effects for the different spin states.
The numeric value of polarization $P_\Lambda$ can be large since there are
no small factors in (\ref{polllg}). We use the model \cite{prdflat} where $M$ is
proportional to two nucleon masses, the value of parameter $\zeta \simeq 2$ and
$r_{U\to S}^{flip}\simeq
0.1-0.2$ fm. The above qualitative
 features of polarization dependence on $x_F$,
$p_\perp$ and energy are in a good agreement with the experimentally observed trends\cite{newrev}.
For example, Fig. 2 demonstrates that the linear $x_F$ dependence is in a good agreement with
the experimental data in the fragmentation region ($x_F\geq 0.4$) where the model
should work. Of course,
the conclusion on $p_\perp$--independence of $\Lambda$- polarization is a rather approximate one
and insignificant deviations from such behavior cannot be excluded.
 \begin{figure}[htb]
\begin{center}
  \resizebox{4cm}{!}{\includegraphics*{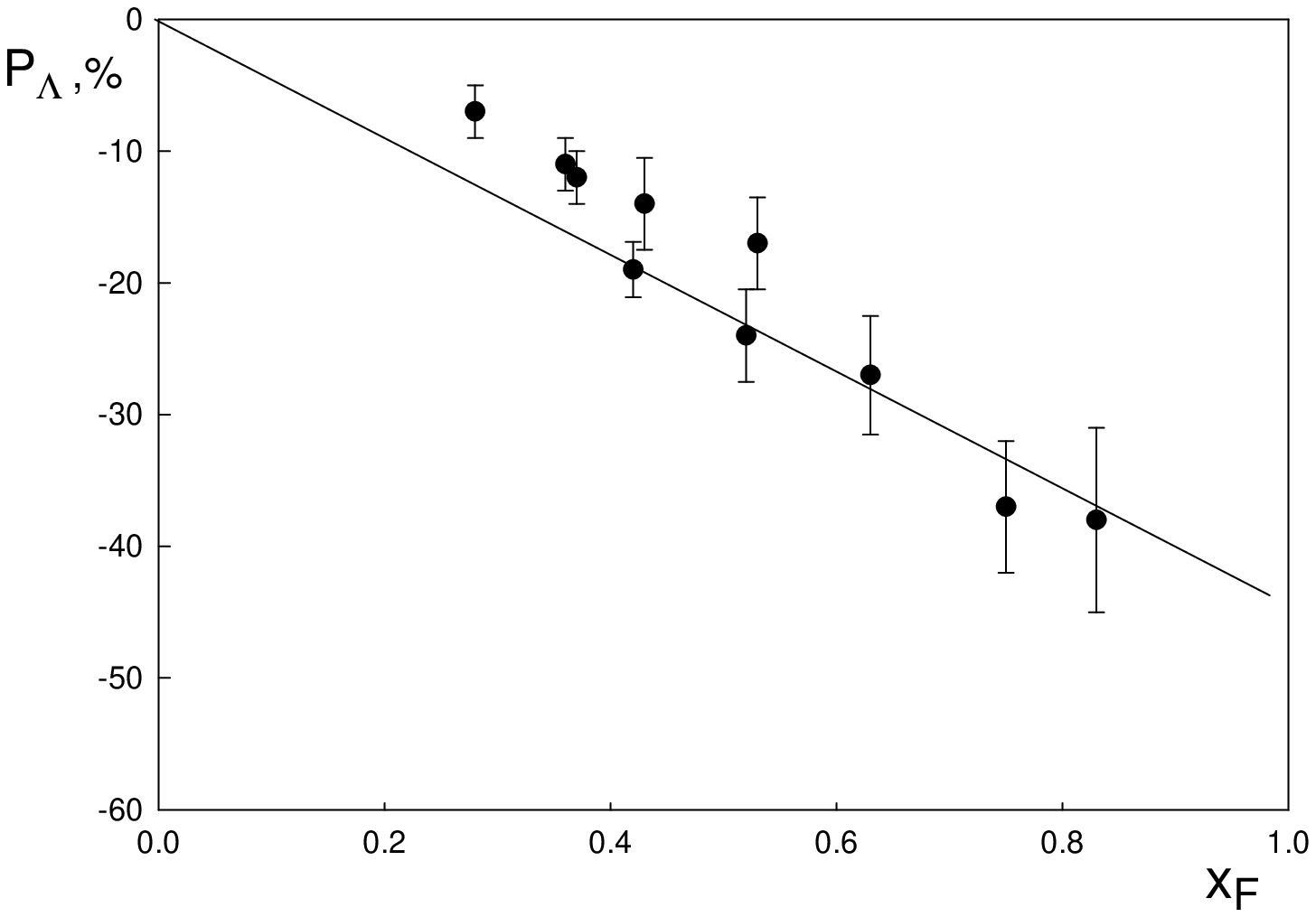}}\;\;\quad
  \resizebox{4cm}{!}{\includegraphics*{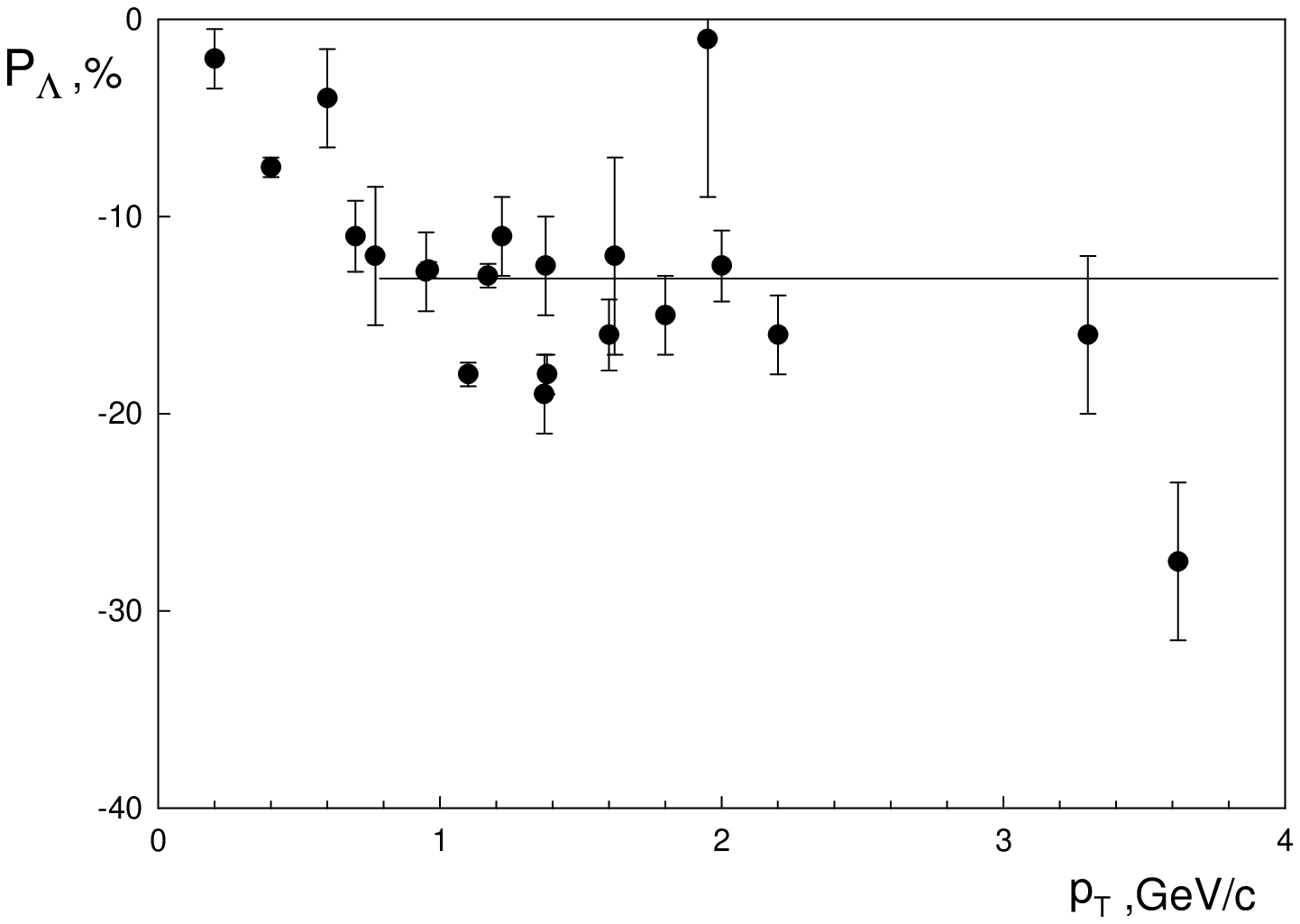}}
\end{center}
\caption{$x_F$ (left panel) and $p_T$ (right panel)
 dependencies of the $\Lambda$-hyperon
polarization} \label{ts}
\end{figure}

The proposed mechanism deals with effective degrees of freedom and takes into
account collective aspects of QCD dynamics. Together with unitarity, which is an essential
ingredient of this approach, it allows  to
 obtained results for polarization dependence on kinematical variables
 in  agreement with the  experimental  behavior
of $\Lambda$-hyperon polarization, i.e.
 linear dependence on $x_F $ and
flat dependence
on $p_\perp$ at large $p_\perp$
in the fragmentation region are reproduced.
Those dependencies together with the energy independent
behavior of polarization at large transverse
momenta are the straightforward consequences of this model.

We discussed  polarization in the particle production in the fragmentation region.
In the central region where correlations
 between impact parameter of the initial and impact parameters of the final particles
 are weak, polarization cannot be generated due to the chiral quark filtering
 mechanism. It is also valid for QGP production as a result of the vacuum excitation. The transverse polarization of $\Lambda$ is expected to be zero too \cite{newrev}. Thus, the detection of vanishing transverse polarization might be associated with QGP production. Of course, zeroing of transverse polarization is not sufficient to conclude on QGP formation. It is important to note in view of the recent results of the ALICE measurements \cite{alice}, where the data show for the first time that the yields of strangeness increases with multiplicity compared  to the yield of pions. Such enhancement at high multiplicities could be interpreted as a signal of QCP formation in small systems (cf. \cite{alice} and references therein).
 
 The measurements performed by ATLAS Collaboration \cite{atlas} at the LHC are  also in favor of the above conclusion.
Moreover, it is  clear that since antiquarks are produced through spin-zero Goldstone bosons
we should expect transverse polarization $P_{\bar\Lambda}\simeq 0$.
The chiral quark filtering is also relatively suppressed when compared to direct elastic
 scattering of quarks and therefore
   should not play a role in the reaction $pp\to pX$ in the fragmentation
 region, i.e. protons should be produced unpolarized. Indeed, these features take place
 in the experimental data set.

We considered here the mechanism leading to polarization of $\Lambda$ resulting from fragmentation of  a colliding proton. 
From this point of view it seems  rather naturally to expect a large polarization of $\Lambda$ in the process of diffraction dissociation
\[
p+p\to \Lambda +K^+ +X.
\]
The measurements performed at ISR  \cite{erhan} are in agreement with this observation.
\section{On the $\Lambda$-polarization measurements at the LHC}
The results discussed in the previous section demonstrate that studies of spin effects at such high energy
 as the LHC provides are not senseless. Here we mention a simple experimental feasibility for performing 
 those measurements. The measurements are based on the studies of the weak decay of $\Lambda$ into $p$ and $\pi^-$ which
 allows one to reconstruct $P_\Lambda$ from the angular distribution of the proton $dN/d \cos \theta_p$ produced as a result of this decay.
 It can be performed since this angular distribution is proportional to
 \[
 1+\alpha_\Lambda P_\Lambda \cos \theta_p,
 \]
where  $\theta_p$ is the angle between the nucleon momentum and the axis of the hyperon's polarization. Plotting the distribution $dN/d \cos \theta_p$ against $\cos \theta_p$ the polarization can be obtained since the value of the decay parameter $\alpha_\Lambda$ is known. Two-track events should be used to reconstruct $\Lambda\to p \pi^-$ weak decays and it seems promising to use T1 and T2  inelastic telescopes of the  
TOTEM experiment as a base for the relevant experimental set-up.  

The $\Lambda$ polarization direction is parallel to the normal to the production plane $\hat{\vec{n}}$. This is the result of the parity conservation in strong interactions.
The unit vector $\hat{\vec{n}}$ is determined by the  product
\[
\hat{\vec{n}}=\frac{\vec{p}_b\ \times \vec{p}_\Lambda}{|\vec{p}_b\ \times \vec{p}_\Lambda|}
\]
of the beam momentum $\vec{p}_b$ and $\vec{p}_\Lambda$ (momentum of $\Lambda$) . In central rapidity region, the transverse  polarization of $\Lambda$ has been measured at the LHC \cite{atlas}, where small value of it has been found.

However, polarization of $\Lambda$ in the diffraction dissociation processes is expected to be at the level 30-40\% on the base of the experimental data extrapolations and simple semiclassical mechanisms' estimations. It is important also to check an energy independence of the hyperon polarization observed at lower energies to bring deeper understanding of the diffractive physics and its  dependence on spin.

One could  note that the instrumental experience  obtained at CERN ISR at polarization measurements of $\Lambda$ (under the use of e.g. R608 forward spectrometer \cite{erhan}) could be helpful at the LHC too, and especially in the diffractive dissociation processes.

The measurements discussed above are also important for the QGP detection. The importance of the transverse polarization measurements has already been noted. 
An indirect way to measure intensity of the multistrange baryon production has been earlier discussed in \cite{newrev} and is based on the studies of the longitudinal polarization of  $\bar{\Lambda}$ produced in the weak decay $\bar{\Xi}\to\bar{\Lambda}+\pi$ of the unpolarized $\bar{\Xi}$.  This polarization arises due to parity nonconservation at this weak decay process. As it was mentioned in \cite{newrev}, the QGP formation might lead to a very significant longitudinal polarization of  $\bar{\Lambda}$.

Finally, spin studies at the LHC are a relevant tool for the strong sector of the Standard Model test as well as these could serve one of the important probes of the QGP formation

\end{document}